\documentclass[american,prl, twocolumn]{revtex4-1}
\usepackage[T1]{fontenc}
\usepackage[utf8]{inputenc}
\usepackage{babel}
\usepackage{array}
\usepackage{float}
\usepackage{multirow}
\usepackage{amsmath}
\usepackage{amssymb}
\usepackage{graphicx}
\usepackage{setspace}
\usepackage[unicode=true]
 {hyperref}

\makeatletter

\providecommand{\tabularnewline}{\\}
\newcommand{\lyxdot}{.}

\usepackage{latexsym}
\usepackage{textgreek}
\@ifundefined{definecolor}
 {\usepackage{color}}{}
\usepackage[table]{xcolor}
\usepackage{graphicx}
\usepackage{hyperref} 
\usepackage{url}

\usepackage{lineno}

\newcommand*\patchAmsMathEnvironmentForLineno[1]{%
  \expandafter\let\csname old#1\expandafter\endcsname\csname #1\endcsname
  \expandafter\let\csname oldend#1\expandafter\endcsname\csname end#1\endcsname
  \renewenvironment{#1}%
     {\linenomath\csname old#1\endcsname}%
     {\csname oldend#1\endcsname\endlinenomath}}%
\newcommand*\patchBothAmsMathEnvironmentsForLineno[1]{%
  \patchAmsMathEnvironmentForLineno{#1}%
  \patchAmsMathEnvironmentForLineno{#1*}}%
\AtBeginDocument{%
\patchBothAmsMathEnvironmentsForLineno{equation}%
\patchBothAmsMathEnvironmentsForLineno{align}%
\patchBothAmsMathEnvironmentsForLineno{flalign}%
\patchBothAmsMathEnvironmentsForLineno{alignat}%
\patchBothAmsMathEnvironmentsForLineno{gather}%
\patchBothAmsMathEnvironmentsForLineno{multline}%
}

\newcommand{\mv}[1]{\mathbf{#1}}

\newcommand{\width}{\lambda}
\newcommand{\intEnergy}{\Gamma}
\newcommand{\kin}{M}
\newcommand{\dx}{\Delta x}
\newcommand{\dt}{\Delta t}
\newcommand{\df}{\mu}
\newcommand{\interp}{h}
\newcommand{\gridCoup}{a} 
\newcommand{\EngFunc}{F}
\newcommand{\engDens}{f}
\newcommand{\capLength}{d_0}
\newcommand{\ponderation}{\gamma}
\newcommand{\temperature}{U}
\newcommand{\grid}{\mathbf{p}}
\newcommand{\nbs}{j}
\newcommand{\dir}{k}

\newcommand{\iFact}{R}
\newcommand{\gvec}{\mathbf{r}_{\dir}}

\definecolor{link_color}{HTML}{00406E} 
\definecolor{cite_color}{HTML}{590000} 
\definecolor{dark-violet}{HTML}{9400d3} 
\definecolor{forest-green}{HTML}{228b22} 
\definecolor{dark-red}{HTML}{8b0000} 
\definecolor{dark-blue}{HTML}{00008b} 
\definecolor{dark-pink}{HTML}{ff1493}
\definecolor{dark-yellow}{HTML}{c8c800}
\definecolor{dark-orange}{HTML}{c04000}
\definecolor{dark-salmon}{HTML}{e9967a}
\definecolor{midnight-blue}{HTML}{191970}
\definecolor{dark-spring-green}{HTML}{008040}
\definecolor{newcolor}{rgb}{.8,.349,.1}

\hypersetup{
    colorlinks   = true, 
    urlcolor     = link_color, 
    linkcolor    = black, 
    frenchlinks  = true, 
    citecolor    = link_color 
}

\newcommand{\appendixpagenumbering}{
  \break
  \pagenumbering{arabic}
  \renewcommand{\thepage}{S\arabic{page}}
}

\makeatother

\begin{document}
\title{Sharp phase-field modeling of isotropic solidification with a super
efficient spatial resolution}
\author{Michael Fleck\href{https://orcid.org/0000-0002-2799-9075}{\includegraphics[width=10pt]{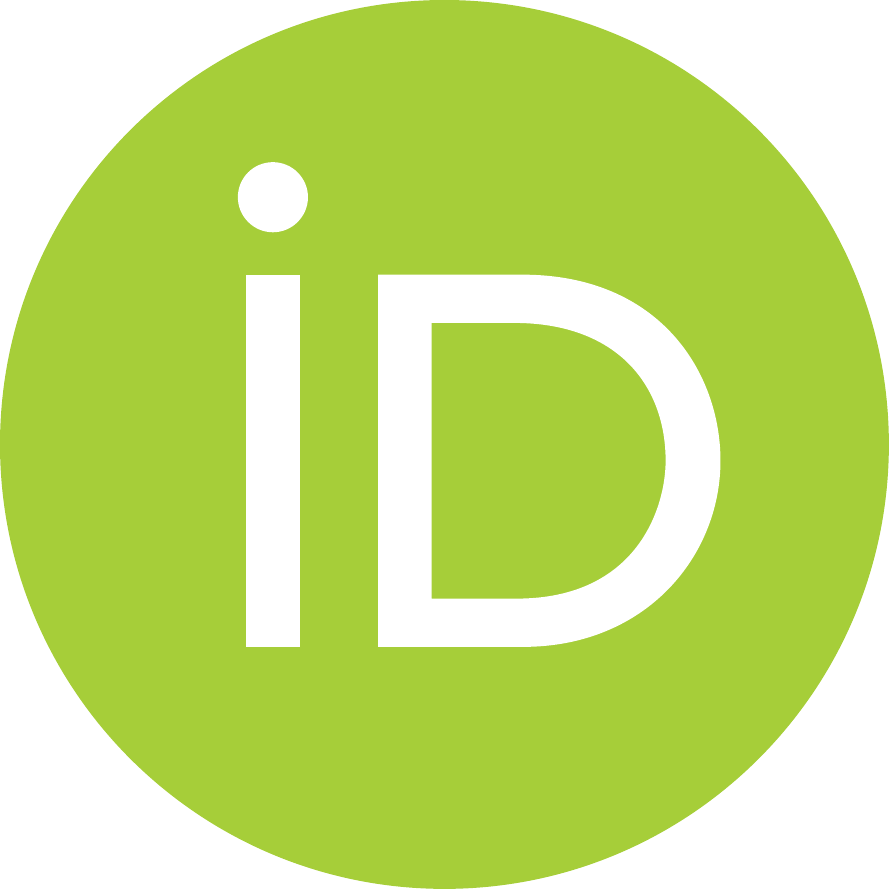}}}
\email{michael.fleck@uni-bayreuth.de}

\affiliation{Metals and Alloys, University of Bayreuth, Prof.-Rüdiger-Bormann-Straße
1, 95447 Bayreuth, Bavaria, Germany}
\author{Felix Schleifer\href{https://orcid.org/0000-0003-1189-9288}{\includegraphics[width=10pt]{./ORCID-iD_icon-vector}}}
\affiliation{Metals and Alloys, University of Bayreuth, Prof.-Rüdiger-Bormann-Straße
1, 95447 Bayreuth, Bavaria, Germany}
\begin{abstract}
The numerical resolution efficiency of phase-field models is limited
by grid friction, grid anisotropy and pinning. The 1D sharp phase-field
model eliminates grid friction and pinning by a global restoration
of Translational Invariance (TI) in the discretized phase-field equation
(Phys.~Rev.~Lett.~121, 025501, 2018). In 3D global TI restricts
the beneficial modeling properties to a finite number of fixed interface
orientations. We propose an accurate scheme to restore TI locally
in the local interface normal direction. At one-grid-point interface
resolutions, the new model captures the formation of isotropic seaweed
structures without spurious dendritic selection by grid anisotropy. 

\end{abstract}
\maketitle
Diffuse interface descriptions, such as phase-field models, are
widely used for the microscopic modeling of solidification as well
as related microstructure evolution problems \citep{KurzRappazTrievedi2021PartII,TourretLiuLLorca2021,TonksAagesen2019}.
 Quantitative simulations require a proper numerical resolution
of the diffuse solid/liquid interface, i.e. the diffuse interface
profile has to be resolved by a certain minimal amount of grid points.
In case of numerical under-resolution, the simulation is subjected
to spurious grid anisotropy as well as grid friction, which in the
worst case leads to the ``pinning'' of the diffuse interface on
the computational grid. In conventional phase-field models the minimal
number of grid points used to resolve the profile is about $4$ \citep{JokisaariVoorheesGuyerWarrenHeinonen2018}.
In our notation this corresponds to the dimensionless grid resolution
number $\tilde{\width}=2$. However, depending on the accuracy demands
of the simulation, the double, triple or even quadruple amount of
grid points can be required. 

Recently, Finel et al.~found a striking new way to eliminate grid
friction and pinning in one dimension, called the sharp phase-field
model \citep{FinelLeBouarDabasAppolairYamada2018}. This method is
conceptually related to other techniques to improve the performance
of phase-field models based on the phase-field profile function \citep{Glasner2001,Weiser2009,Eiken2012,DebierreGuerinKassner2016,ShenXuYang2019,JIMolaviTabriziKarma2022}.
The 1D sharp phase-field model operates at one-grid-point profile
resolutions ($\tilde{\width}=0.5$) and below without the occurrence
of grid pinning! 

However, beside the profile resolution, there is one other important
aspect that limits the spatial resolution efficiency of phase-field
models in general: They cannot operate at arbitrarily small interface
energy densities $\intEnergy$. Consider an interface between two
phases at different bulk free energy density levels. The latter, also
called the driving force $\df$, induces an interface motion lowering
the total free energy of the system. For too small interface energies
or too large driving forces either the high energy phase turns unstable
(phase stability limit) or the phase-field profile is spuriously altered.
The alternation is accompanied by strong grid friction effects. We
define the dimensionless driving force $\tilde{\df}=\df\dx/\intEnergy$,
which relates to the spatial resolution of the simulation via the
grid spacing $\dx$. Imposing constant driving forces, we consider
the simulation of stationary interface motion in 1D at different dimensionless
spatial resolution numbers $\tilde{\width},\tilde{\df}$. Reasonable
model operation at the resolution $\tilde{\width},\tilde{\df}$ is
said to require phase stability and less than $10\%$ relative deviations
from the energetically exact interface velocity. Further details on
this study are given in the supplementary material. In Fig.~\ref{fig:Comparison-of-the-parameter-ranges},
we compare the resulting parameter windows of reasonable model operation
for the most frequently used conventional phase-field model (blue)
and the sharp phase-field model (green). The elimination of spurious
grid friction in the sharp phase-field model allows for orders of
magnitude more efficient simulations than possible with the conventional
phase-field model. 

\begin{figure}[b]
\begin{centering}
{\scriptsize{}}\begin{center}
{\footnotesize{}\input{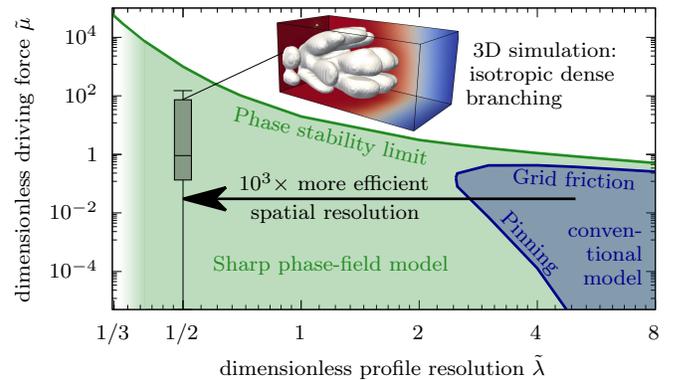}}{\footnotesize\par}
\par\end{center}

{\footnotesize{}\vspace{-12pt}
}{\footnotesize\par}{\scriptsize\par}
\par\end{centering}
\caption{\label{fig:Comparison-of-the-parameter-ranges}Comparison of the parameter
windows of reasonable model operation (stationary interface motion
with relative errors $<0.1$) for two different phase-field models:
The most frequently used conventional model (blue area) and the new
sharp phase-field model (green area).}
\end{figure}

 During diffusion limited solidification the complex evolution of
the solid/liquid interface undergoes a branching instability \citep{MullinsSekerka1964}.
In a fully isotropic system, this leads to the self-organized formation
of so-called isotropic dense branching or seaweed microstructures
\citep{BrenerHMKTemkinAbel1998}, as visible in the inset of Fig.~\ref{fig:Comparison-of-the-parameter-ranges}.
The structure shows a characteristic distance between branches, which
nontrivially relates to the atomistically small capillary length $\capLength$,
that is proportional to the interface energy density $\intEnergy$
\citep{KurzRappazTrievedi2021PartII}. A fundamental challenge in
solidification modeling is the fact that the microscopic distance
between branches is typically several orders of magnitude larger than
a central aspect of its cause, i.e.~the atomistically small capillary
length. If, however, the phase-field model is able to stably operate
at a certain small interface energy or, in other words, a certain
large dimensionless driving force, then the grid spacing $\dx$ can
exceed $\capLength$ in a respective proportion \citep{FleckQuerfurthGlatzel2017}.

Here, we propose a new sharp phase-field model, which captures the
3D formation of isotropic dense branching even at one-grid-point profile
resolutions ($\tilde{\width}=0.5$), see Fig.~\ref{fig:Comparison-of-the-parameter-ranges}!
The absence of any spurious dendritic selection by the computational
grid indicates quite high degrees of isotropy \citep{Ihle2000,BragardKarmaLee2002}.
In this simulation the driving forces are largely inhomogeneous. In
Fig.~\ref{fig:Comparison-of-the-parameter-ranges}, we visualize
the respective driving force distribution by a boxplot with whiskers
to the maximal and minimal value. In this work, we show that the sharp
phase-field model provides quantitative interface velocities within
the full range of different driving forces! To achieve a comparable
accuracy over a similarly wide range of driving forces, the conventional
phase-field model would require profile resolutions of $\tilde{\width}=5$,
as shown in Fig.~\ref{fig:Comparison-of-the-parameter-ranges}. In
this regard, the new sharp phase-field model allows for 3D simulations
of isotropic solidification with a $10^{3}\times$more efficient spatial
resolution.

\paragraph{The new sharp phase-field model}

\label{sec:Sharp-Phase-Field-Modeling}The derivation of the new sharp
phase-field formulation is started from a discrete Helmholtz free
energy functional $F\left[\phi_{\grid}\right]=\sum_{\grid}\engDens_{\grid}\dx^{3}$,
where $\grid$ denotes the locations of the grid points within the
simple cubic 3D numerical lattice with a grid spacing $\dx$. The
discrete Helmholtz free energy density $\engDens_{\grid}$ associated
with the grid point $\grid$ is
\begin{align}
\engDens_{\grid}\!= & \frac{\intEnergy}{C_{\intEnergy}\width}\sum_{\nbs,\dir}\!\ponderation_{\nbs}\nu_{\nbs}\Big(\frac{\width^{2}}{2}(\partial_{\dir}^{+}\phi_{\grid})^{2}\!+\!g_{\dir}(\phi_{\grid})\Big)\!+\!\df_{\grid}\interp(\phi_{\grid}).\label{eq:SPFM-Interface-energy-density-3D}
\end{align}
We restrict the interaction between grid points to the first three
neighboring shells $\nbs\!=\!1,2,3$, with $\left|\gvec\right|_{\nbs}\!=\!\sqrt{\nbs}\dx$
and $\gvec$ being a numerical lattice vector that connects two neighboring
grid points along the direction $\dir$. $\partial_{\dir}^{+}\phi_{\grid}$
denotes the discrete directional derivative, which is approximated
by the forward finite difference expression $\partial_{\dir}^{+}\phi_{\grid}\!\equiv\!(\phi_{\grid+\gvec}\!-\!\phi_{\grid})/\left|\gvec\right|$.
For a given neighboring shell with $m_{\nbs}$ neighboring nodes,
the coefficients $\nu_{\nbs}\!=\!3/m_{\nbs}$ correct for the multiplicity
of the shell. Similar to \citep{FinelLeBouarDabasAppolairYamada2018},
the ponderation coefficients $\ponderation_{\nbs}$ are chosen to
get best possible energetic equality of differently oriented ideal
interfaces. 

The equilibrium potentials $g_{\dir}(\phi)$ are minimal at $\phi\!=\!0$
and $\phi\!=\!1$, which corresponds to the two distinct phases of
the system. $\width$ denotes the width of the diffuse interface,
$\intEnergy$ is the interface energy density, and $C_{\intEnergy}$
is the interface energy calibration parameter. A positive bulk free
energy density difference $\df_{\grid}$ favors the growth of phase
$\phi\!=\!0$ on the expanse of phase $\phi\!=\!1$. Concerning the
interpolation function $\interp(\phi)$, we focus on the natural interpolation
$\interp_{3}\!=\!\phi^{2}(3\!-\!2\phi)$ \citep{FinelLeBouarDabasAppolairYamada2018}
and the most frequently used polynomial $\interp_{\mathrm{5}}\!=\!\phi^{3}(10\!-\!15\phi\!+\!6\phi^{2})$
\citep{Plapp092011,OhnoTakakiShibuta2017,AagesenGaoSchwenAhmed2018,GreenwoodShampurOforiOpokuPinomaaGurevichProvatas2018,GranasyTothWarrenTegzeRatkaiPusztai2019,KimShermanAagesenVoorhees2020}.

The functional phase-field derivative of the discrete Helmholtz free
energy is given by $\delta_{\phi}F\!=\!\partial_{\phi}f_{\grid}\!-\!\sum_{\nbs,\dir}\partial_{\dir}^{-}(\partial_{\left(\partial_{\dir}\phi\right)}^{+}f_{\grid}),$
where the second directional derivative $\partial_{\dir}^{-}$ is
approximated by $\partial_{\dir}^{-}\left(\partial f_{\grid}\right)\!\equiv\!\left(\partial f_{\grid}\!-\!\partial f_{\grid-\gvec}\right)/\left|\gvec\right|$.
The phase-field evolution equation demands that the time derivative
$\partial_{t}\phi_{\grid}$ is proportional to $-\delta_{\phi}\EngFunc$.
We write, $3\width\intEnergy\partial_{t}\phi_{\grid}\!=\!-2\kin\delta_{\phi}\EngFunc$,
where $\kin$ is a kinetic coefficient with the dimension $\left[\kin\right]\!=\!\mathrm{m}^{2}\mathrm{s}^{-1}$\citep{FleckFedermannPogorelov2018}.
During stationary interface motion, driven by a constant $\df$, total
energy conservation demands $v_{\mathrm{th}}\!=\!-\kin\df/\intEnergy$.
The phase-field profile function is 
\begin{align}
\phi_{\mathrm{\grid}} & =\left(1-\tanh2\left(\grid\cdot\mv{n}-c_{n}\right)/\width\right)/2,\label{eq:phase-field-tanh-profile-solution}
\end{align}
which is an analytic solution of the continuum phase-field equation,
if $g(\phi)\!=\!\sum_{\nbs,\dir}g_{\dir}\!\equiv\!8\phi^{2}\left(1\!-\!\phi\right)^{2}$and
$\interp(\phi)\!=\!\interp_{3}\!=\!\phi^{2}(3\!-\!2\phi)$. $\mv{n}$
is the unit normal interface vector and $c_{n}\!=\!v_{\mathrm{th}}t$
denotes the central interface position, moving with the velocity $v_{\mathrm{th}}$.
The profile width of $2\width$ is understood as $96.4\%$ of the
total transition from $\phi\!=\!0$ to $\phi\!=\!1$ ($\tanh2\!\simeq\!0.964$)
\citep{DimokratiLeBouarBenyoucefFinel2020}. 

For vanishing driving forces $\df\!=\!0$ and no phase-field motion
$\partial_{t}\phi\!=\!0$ and the phase-field equation reduces to
\begin{align}
{\textstyle \sum_{\nbs,\dir}}\ponderation_{\nbs}\nu_{\nbs}\{\width^{2}(\phi_{\grid+\gvec}\!\!-\!2\phi_{\grid}\!+\!\phi_{\grid-\gvec})/\gvec^{2}-\partial_{\phi}g_{\dir}\} & =0,\label{eq:discrete-equilibirum-formulation}
\end{align}
where $\partial_{\phi}\!=\!\partial/\partial\phi$ denotes the partial
phase-field derivative. The condition holds, if all individual $\dir-$components
are simultaneously satisfied. Those can be satisfied at any real time
during the propagation of the interface using the addition property
of the hyperbolic tangent profile (\ref{eq:phase-field-tanh-profile-solution})
$\phi_{\grid\pm\gvec}\!=\!(1\!\pm\!\gridCoup_{\dir})\phi_{\grid}/(1\!\pm\!(2\phi_{\grid}\!-\!1)\gridCoup_{\dir}),$
where the grid coupling parameters $\gridCoup_{\dir}\!\left(\mv{n}\right)$
are defined as $\gridCoup_{\dir}\!=\!\tanh\left(2\gvec\!\cdot\!\mv{n}/\width\right).$
Inserting this property into the phase-field equilibrium condition,
we obtain the $\dir-$th component of the modified equilibrium potential
\begin{align}
g_{\dir}(\phi)\frac{\gvec^{2}}{\width^{2}}=\phi(1\!-\!\phi)\,+\, & \frac{1\!-\!\gridCoup_{\dir}^{2}}{4\gridCoup_{\dir}^{2}}\ln\bigg(\frac{1\!-\!\gridCoup_{\dir}^{2}}{1\!-\!\gridCoup_{\dir}^{2}\left(1\!-\!2\phi\right)^{2}}\bigg),\label{eq:equilibrium-potential-TI}
\end{align}
which further satisfies $g_{\dir}(\phi\!=\!0,\!1)\!=\!0$, to allow
an easy calculation of the system's total interface energy by $F_{\mathrm{int}}(\phi_{\grid})\!=\!\sum_{\grid}\engDens_{\df=0}$
using an arbitrary phase-field \citep{SchleiferHolzingerLinGlatzelFleck2019,SchleiferFleckHolzingerLinGlatzel_superalloys2020}.
In the continuum limit $\left|\gvec\right|\!\rightarrow\!0$, Eq.~(\ref{eq:equilibrium-potential-TI})
converges to the conventional Continuum Field (CF) potential $g_{\dir}^{\infty}\!=\!8\phi^{2}\left(1\!-\!\phi\right)^{2}$.

\begin{figure}
\begin{centering}
\begin{minipage}[t]{9cm}%
\begin{center}
\includegraphics[width=8cm]{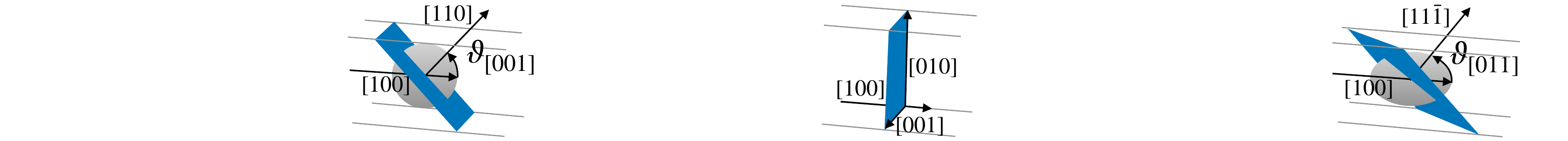}
\par\end{center}
\begin{flushleft}
{\footnotesize{}}\begin{flushleft}
{\footnotesize{}\vspace{-10pt}
\input{./pinning_force_vs_orientation.dtex}}
\par\end{flushleft}{\footnotesize\par}
\par\end{flushleft}%
\end{minipage}
\par\end{centering}
\caption{\label{fig:Orientation-model}Test of Translational Invariance (TI)
of the ideal profile (\ref{eq:phase-field-tanh-profile-solution})
within the equilibrium condition (\ref{eq:discrete-equilibirum-formulation}).
We plot the oscillation amplitude $A$ of the system integral over
Eq.~(\ref{eq:discrete-equilibirum-formulation}) during profile motion
for different interface orientation angles $\vartheta_{[001]}$ and
$\vartheta_{[011]}$. Profile resolution $\tilde{\width}\!=\!\width/\dx\!=\!0.5$;
system size $300\!\times\!1\!\times\!1$.}
\end{figure}

Translational Invariance (TI) in the phase-field equation is restored
based on properties of the profile function (\ref{eq:phase-field-tanh-profile-solution}).
Without TI, the system integral over Eq.~(\ref{eq:discrete-equilibirum-formulation})
oscillates, when the ideal profile (\ref{eq:phase-field-tanh-profile-solution})
is moved on the grid, as plotted in Fig.~\ref{fig:Orientation-model}.
To determine the grid coupling parameters $\gridCoup_{\dir}\!\left(\mv{n}\right)$
Finel et al.~proposed to represent the interface normal vector $\mv{n}$
by a constant unit vector $\mv{u}$, perpendicular to a properly chosen
lattice plane \citep{FinelLeBouarDabasAppolairYamada2018}.  This
globally restores TI for interface orientations that agree to one
of the equivalent lattice orientations $\left\langle \mv{u}\right\rangle $,
as shown in Fig.~\ref{fig:Orientation-model} for different TI$_{\left\langle \mv{u}\right\rangle }$-models.
Further details are given in the supplementary material. The newly
proposed TI$_{\left\langle \mv{n}\right\rangle }$-model (green curve)
uses grid coupling parameters calculated from the local interface
normal direction, leading to very small oscillations regardless of
the interface orientation. 

For the sufficiently accurate determination of $\gridCoup_{\dir}\!\left(\mv{n}\right)$,
we proceed as follows: First, we calculate preliminary grid coupling
parameters by $\hat{\gridCoup}_{\dir}\!=\!(\hat{\gridCoup}_{\dir}^{+}\!+\!\hat{\gridCoup}_{\dir}^{-})/2$,
where
\begin{align}
\hat{\gridCoup}_{\dir}^{\pm} & =\frac{\pm\left(\phi_{\grid\pm\gvec}-\phi_{\grid}\right)}{\phi_{\grid}-2\phi_{\grid\pm\gvec}\phi_{\grid}+\phi_{\grid\pm\gvec}}.\label{eq:grid-coupling-calculation-scheme}
\end{align}
Using the modified equilibrium potentials the explicit dependence
of the phase-field equation on the profile width $\width$ cancels
out. Then, $\width$ is solely controlled by the preliminary grid
coupling parameters, which also contain the a priori unknown interface
normal vector $\hat{\mv{n}}$. Thus, without length control of $\hat{\mv{n}}$
the profile width $\width$ wouldn't be defined in the model. Thus,
we locally calculate all components of the interface normal vector
$\hat{n}_{\dir}\!=\!\width\mathrm{arctanh}(\hat{\gridCoup}_{\dir})/\left|2\gvec\right|$,
restore unit length via $\mv{n}\!=\!\hat{\mv{n}}/\left|\hat{\mv{n}}\right|$
and calculate corrected grid coupling parameters $\gridCoup_{\dir}\left(\mv{n}\right)$. 

The advancing solidification is accompanied by a release of latent
heat at the solid/liquid interface \citep{KassnerGuerinDucouss082010}.
Thus, the dimensionless temperature field $\temperature_{\grid}\!=\!C(T_{\grid}\!-\!T_{M})/L$
is introduced, where $T_{M}$, $L$ and $C$ denote the melting temperature,
latent heat and heat capacity, respectively \citep{FleckHuterPilipen012010}.
The driving force for solidification is given by $\df_{\grid}\!=\!-\temperature_{\grid}\intEnergy/\capLength$,
where $\capLength\!=\!\intEnergy T_{M}C/L^{2}$ denotes the capillary
length. The temperature obeys a diffusion equation, $\partial_{t}\temperature_{\grid}\!=\!D\nabla^{2}\temperature_{\grid}\!+\!\iFact(\phi_{\grid})\partial_{\phi}\interp\partial_{t}\phi_{\grid}$,
with equal diffusion coefficients $D$ in the solid and liquid phase.
For small phase-field widths, $\width/\dx\!\leq\!2$, and $\iFact\!=\!1$,
we observe spuriously inhomogeneous releases of latent heat, whenever
a grid point is close to the interface center. The spurious heat release
provides oscillations in the solidification velocity as well as some
degree of kinetic anisotropy. Therefore, we propose the regularization
$\iFact(\phi_{\grid})$ in the diffusion equation 
\begin{align}
\iFact(\phi_{\grid}) & =\frac{3C_{\iFact}}{4}\frac{\gridCoup_{\left\langle 100\right\rangle }\width}{\dx}\left(1-\gridCoup_{\left\langle 100\right\rangle }^{2}\left(1-2\phi_{\grid}\right)^{2}\right)^{-2},\label{eq:SPFM-definition-iFact}
\end{align}
where the grid coupling parameter is $\gridCoup_{\left\langle 100\right\rangle }\!=\!\tanh2\dx/\width$
and $C_{\iFact}$ denotes a calibration constant, which is required
to maintain total energy conservation during solidification. The dependence
of $C_{\iFact}$ as a function of the dimensionless profile resolution
is plotted in Fig.~\ref{fig:energy-Calibration}.

\paragraph{Model calibration}

 The interface energy calibration $C_{\intEnergy}$ is calculated
via $C_{\intEnergy}\!=\!\sum_{\grid_{[100]}}\!\mv{e}_{[100]}\!\cdot\!\mv{n}\engDens(\phi_{\grid}(\mv{n}))_{\df=0}/\intEnergy$,
where $\mv{e}_{[100]}$ denotes a unit vector pointing in the $[100]-$direction
of the computational grid, $\sum_{\grid_{[100]}}$ denotes the sum
in the $[100]-$direction, $\mv{n}$ is again the direction normal
to the interface, and the phase-field values $\phi_{\grid}(\mv{n})$
are given by the ideal profile~(\ref{eq:phase-field-tanh-profile-solution})
with orientation $\mv{n}$. For the determination of $C_{\intEnergy}$
we chose the $[100]-$direction as interface orientation. The determination
of the energy calibration factor is independent from the choice of
the ponderation coefficients. Fig.~\ref{fig:energy-Calibration}
shows the phase-field width dependence of the different calibration
factors. The continuum limit for the calibration factor, $C_{\intEnergy}^{\infty}\!=\!2/3$,
is indicated by the solid black line in Fig.~\ref{fig:energy-Calibration}.
For sharp diffuse interfaces with a phase-field width below $\tilde{\width}\!<\!2$,
we obtain substantially smaller values for the calibration line integral
as compared to the limiting value. 
\begin{figure}
\begin{centering}
{\scriptsize{}}\begin{center}
{\footnotesize{}\input{./energy_calibration.dtex}\vspace{-12pt}
}
\par\end{center}{\scriptsize\par}
\par\end{centering}
\caption{\label{fig:energy-Calibration} Plot of the different calibration
parameters $C_{\intEnergy}$ (solid green), $C_{\intEnergy}^{\mathrm{CF}}$
(dashed green), $C_{\iFact}$ (red) and the ponderation coefficients
$\ponderation_{2}$ (violet) and $\ponderation_{3}$ (blue) as a function
of the phase-field width $\tilde{\width}$. $\ponderation_{1}\!=\!1\!-\!\ponderation_{2}\!-\!\ponderation_{3}$}
\end{figure}

For the determination of the ponderation coefficients an optimization
procedure similar to the one proposed by Finel et al.~\citep{FinelLeBouarDabasAppolairYamada2018}
has been developed.  The ponderation coefficients $\ponderation_{\nbs}$
should be chosen such that the interface energy becomes as isotropic
as possible, i.e.~the discrete interface energy integral $\intEnergy(\mv{n})\!=\!\big<\sum_{\grid_{[100]}}\!\mv{e}_{[100]}\!\cdot\!\mv{n}\engDens(\phi_{\grid}(\mv{n}))_{\df=0}\big>_{r_{n}}$
should dependent on the interface orientation $\mv{n}$ as little
as possible. Since at least some of these line integrals may not be
Translationally Invariant, we further average over a number of different
values obtained for different positions $r_{n}$ of the interface
center, as denoted by the angle brackets with index $r_{n}$. Given
a starting set for the ponderation coefficients $\ponderation_{\nbs}$,
we calculate the following three different interface energy densities:
$\intEnergy([100])\!=\!\intEnergy_{\nbs=1}$, $\intEnergy([110])\!=\!\intEnergy_{\nbs=2}$
and $\intEnergy([111])\!=\!\intEnergy_{\nbs=3}$. As a measure for
interface energy isotropy and as the minimization target, the square
root of the sum of the deviations from the average interface energy
value in square of these three energy densities is chosen. i.e.~
\begin{align}
\left\{ \ponderation_{\nbs}\right\} : & \min\sqrt{{\textstyle \sum_{\nbs}}(\overline{\intEnergy}-\intEnergy_{\nbs})^{2}},\label{eq:Ponderation-Determination-optimization-target}
\end{align}
with $\overline{\intEnergy}\!=\!\sum_{\nbs}\intEnergy_{\nbs}/3$.
The optimal choice for the ponderation coefficients $\left\{ \ponderation_{\nbs}\right\} $,
with respect to this minimization target and under the constraint
$\sum_{\nbs}\ponderation_{\nbs}\!=\!1$, has been calculated by a
simple steepest decent algorithm. In Fig.~\ref{fig:energy-Calibration},
the optimal ponderation coefficients are plotted as function of the
phase-field width for the TI$_{\left\langle \mv{n}\right\rangle }$-model
(solid curves) as well as for the TI$_{\left\langle 100\right\rangle }$-model
(dashed curves). The ponderation coefficients obtained for the CF-model
are nearly identical to those of the TI$_{\left\langle 100\right\rangle }$-model.

\begin{figure}
\begin{centering}
\begin{center}
\input{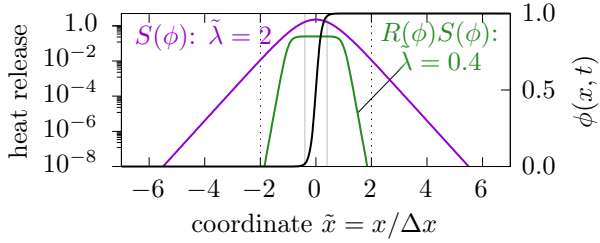}{\scriptsize{}\vspace{-12pt}
}
\par\end{center}
\par\end{centering}
\caption{\label{fig:Source-Term-Calibration}Comparison of the heat release
range of the source term $S(\phi)\!=\!\partial_{\phi}\interp\partial_{t}\phi\!\sim\!\partial_{\phi}\interp^{2}(\phi)$
with ($\tilde{\width}\!=\!0.4$) and without ($\tilde{\width}\!=\!2$)
the regularization factor $\iFact(\phi)$ Eq.~(\ref{eq:SPFM-definition-iFact}).
The comparison is based on the ideal phase-field profile function~(\ref{eq:phase-field-tanh-profile-solution})}
\end{figure}

The idea behind the source term regularization $\iFact(\phi)$ Eq.~(\ref{eq:SPFM-definition-iFact})
is to distribute the latent heat release over a slightly enlarged
range, involving more than just a single grid point. The different
ranges of heat release of different source term variants are compared
in Fig.~\ref{fig:Source-Term-Calibration}. The regularization requires
a phase-field width dependent calibration procedure. For a given phase-field
width the calibration parameter $C_{\iFact}$ has to take a specific
value in order to ensure the conservation of the total energy in the
system. Using some arbitrary starting value for $C_{\iFact}$, we
perform a long term simulations of solidification until quasi two
phase equilibrium in a small, thermally isolated, one-dimensional
system with an initially homogeneous undercooling temperature of $\temperature_{0}\!=\!-0.7$.
Then, based of the deviation of the measured solid phase fraction
from the expected outcome of $0.7$, we successively optimize the
$C_{\iFact}$ value. 

\paragraph{Quantitative stationary solidification}

\begin{figure}
\begin{centering}
{\small{}}{\scriptsize{}}\begin{center}
\begin{minipage}[t][1\totalheight][c]{0.48\textwidth}%
{\small{}\input{./pure-sub-solidi-1D-noIFACT.dtex}}{\small\par}

{\small{}\input{./pure-sub-solidi-1D-comparison.dtex}}{\small\par}%
\end{minipage}
\par\end{center}

\begin{center}
\begin{minipage}[t][1\totalheight][c]{0.48\textwidth}%
{\small{}\input{./pure-sub-solidi-1D-velo.dtex}}{\small\par}

{\footnotesize{}}{\footnotesize\par}%
\end{minipage}{\small{}\vspace{-12pt}
}{\small\par}
\par\end{center}{\scriptsize\par}
\par\end{centering}
\caption{\label{fig:stationary-solidification-study}Stationary solidification
using (i) the Continuum Field model (CF$+\interp_{5}$) for $\tilde{\width}\!=\!2$
in blue, (ii) the Translationally Invariant model (TI$+\interp_{3}$)
for $\tilde{\width}\!=\!0.4$ in red and (iii) the TI-model with regularization
(TI$+\interp_{3}\!+\!\iFact$) in green. a) Exemplary simulation results
and a plot of the velocity as function of the interface center ($\tilde{\df}_{\mathrm{int}}\!=\!100$).
The temperature $\temperature$ is given by colored lines and the
phase-field values by black full symbols. b) Plot of the interface
velocity error as function of the dimensionless driving force $\tilde{\df}_{\mathrm{int}}\!=\!\df_{\mathrm{int}}\dx/\intEnergy$.}
\end{figure}

\label{subsec:Inhomogeneous-driving-forces} In Fig.~\ref{fig:stationary-solidification-study}a),
the configuration of stationary solidification is shown. An animation
of this figure is provided in the supplementary material. Far in front
of the solid/liquid interface the temperature is $\temperature(L)\!=\!-2.0$.
When the system reaches a stationary state, the solid phase is found
at the minimal undercooling temperature of $\temperature_{\mathrm{int}}\!=\!-1.0$.
Then, the theoretically expected solidification velocity is given
by $v_{\mathrm{th}}\!=\!\kin\temperature_{\mathrm{int}}/\capLength,$
where $\kin$ denotes the kinetic coefficient, and $\capLength$ is
the capillary length \citep{Caginalp1989}. We restrict to the comparison
with the sharp interface equation and omit more sophisticated thin
interface corrections \citep{KarmaRappel041998}. The ratio between
the total system length and the theoretic stationary diffusion length
$l_{D}\!=\!2D/v_{\mathrm{th}}$ is chosen to be $L/l_{D}\!=\!5$.
The system is resolved by $200$ grid points, i.e.~$L/\dx\!=\!200$,
with a solid phase fraction of $12\%$. The fraction is kept constant
by incremental shifting of the whole system \citep{FleckBrenerSpatsch2010}.

In Fig.~\ref{fig:stationary-solidification-study}b) the relative
error in the solidification velocity is plotted as function of the
dimensionless driving force $\tilde{\df}_{\mathrm{int}}\!=\!\df_{\mathrm{int}}\dx/\intEnergy$.
The CF$+\interp_{5}$-model (blue color) is subjected to strong spurious
grid friction for both small as well as large dimensionless driving
forces. In case of $\tilde{\df}_{\mathrm{int}}\!=\!100$, the observed
solidification velocity is $90\:\%$ smaller than the expectation.
The TI-models are limited by phase stability only. This limit is
indicated by the vertical dashed line in Fig.~\ref{fig:stationary-solidification-study}b).
The TI$+\interp_{3}$-model (red curve) provides large oscillations
in the interface velocity. These result from spuriously inhomogeneous
heat release at the solid/liquid interface, as visible in Fig.~\ref{fig:stationary-solidification-study}a).
It can be avoided by employing the newly proposed source term regularization
$\iFact$ Eq.~(\ref{eq:SPFM-definition-iFact}), see the green curves
in Fig.~\ref{fig:stationary-solidification-study}. 

\paragraph*{Diffusion limited solidification\label{sec:Diffusion-limited-solidification} }

\begin{figure*}
\centering{}{\small{}}%
\begin{tabular}{c>{\centering}m{2.8cm}ccc}
\hline 
\multirow{5}{*}{{\small{}}%
\begin{minipage}[c]{2.3cm}%
{\small{}\vspace{0.2cm}
initial state\vspace{0.2cm}
\includegraphics[width=0.8\textwidth]{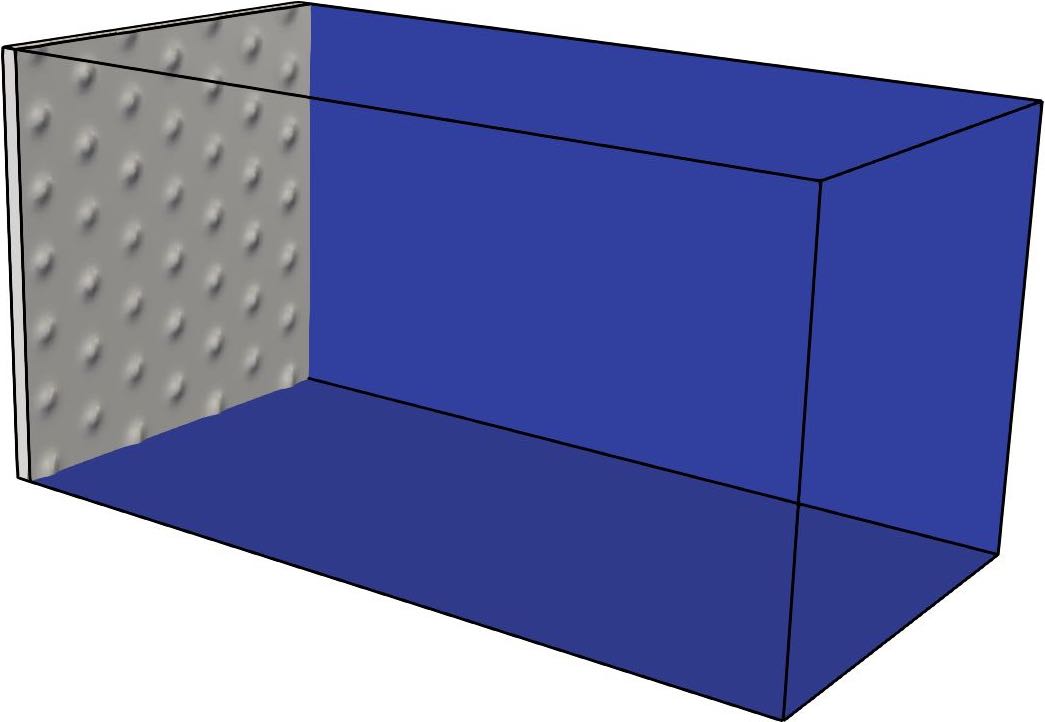}}\\
{\small{}\vspace{0.1cm}
\includegraphics[width=0.6\textwidth]{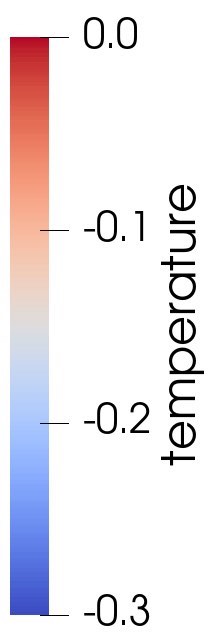}}%
\end{minipage}} & {\small{}time $t\kin/\dx^{2}$:} & {\small{}$5$} & {\small{}$50$} & {\small{}$\geq100$}\tabularnewline
\cline{2-5} \cline{3-5} \cline{4-5} \cline{5-5} 
 & {\small{}TI$_{\left\langle \mv{n}\right\rangle }\!+\!\interp_{3}\!+\!\iFact$}\\
{\small{}$\tilde{\width}=0.5$\vspace{0.5cm}
~} & {\small{}}%
\begin{minipage}[b][1\totalheight][c]{2.7cm}%
{\small{}\vspace{2pt}
\includegraphics[width=0.8\textwidth]{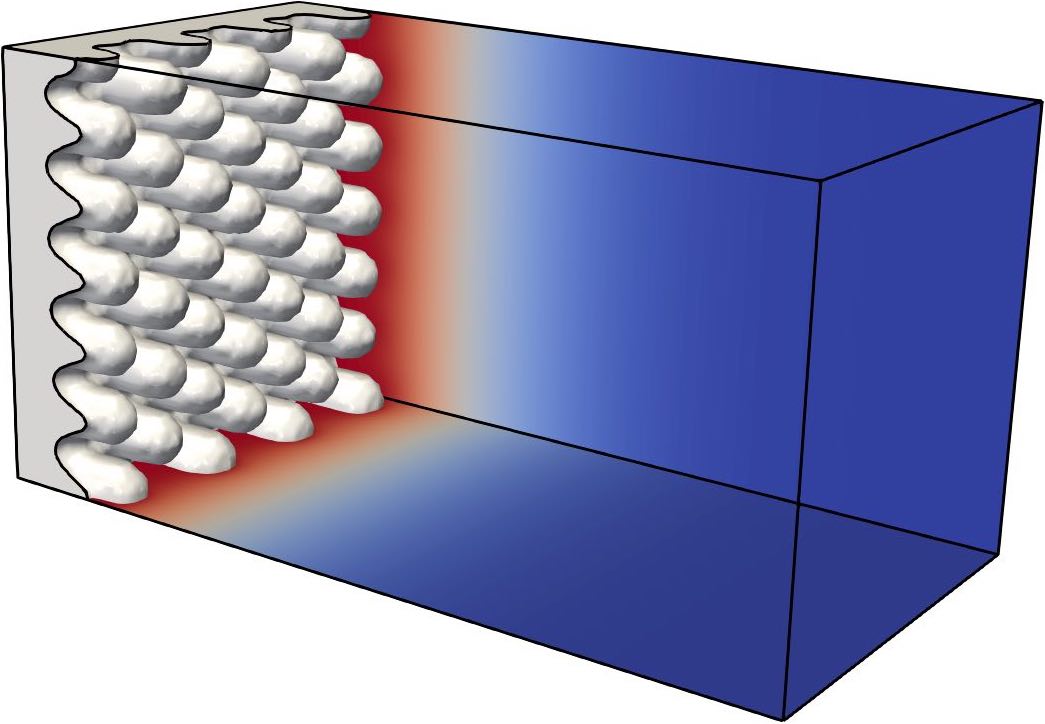}}%
\end{minipage} & {\small{}}%
\begin{minipage}[b][1\totalheight][c]{2.7cm}%
{\small{}\vspace{2pt}
\includegraphics[width=0.8\textwidth]{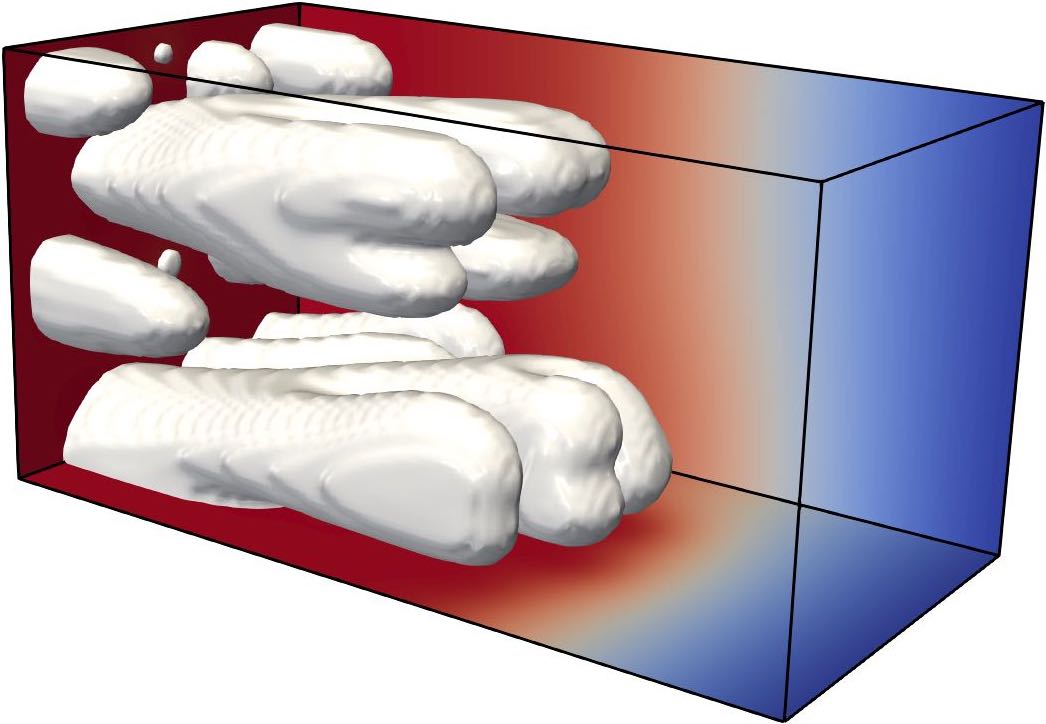}}%
\end{minipage} & {\small{}}%
\begin{minipage}[b][1\totalheight][c]{2.7cm}%
{\small{}\vspace{2pt}
\includegraphics[width=0.8\textwidth]{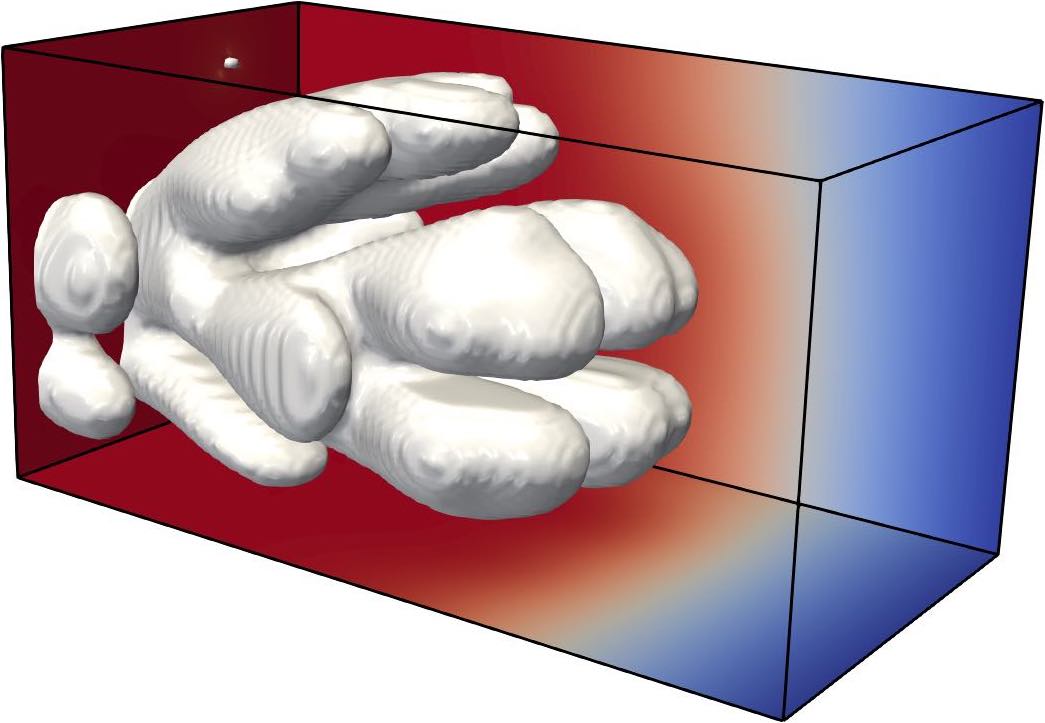}}%
\end{minipage}\tabularnewline
\cline{2-5} \cline{3-5} \cline{4-5} \cline{5-5} 
 & {\small{}TI$_{\left\langle \mv{n}\right\rangle }\!+\!\interp_{3}$}\\
{\small{}$\tilde{\width}=0.5$} & {\small{}}%
\begin{minipage}[c]{2.7cm}%
{\small{}\vspace{2pt}
\includegraphics[width=0.8\textwidth]{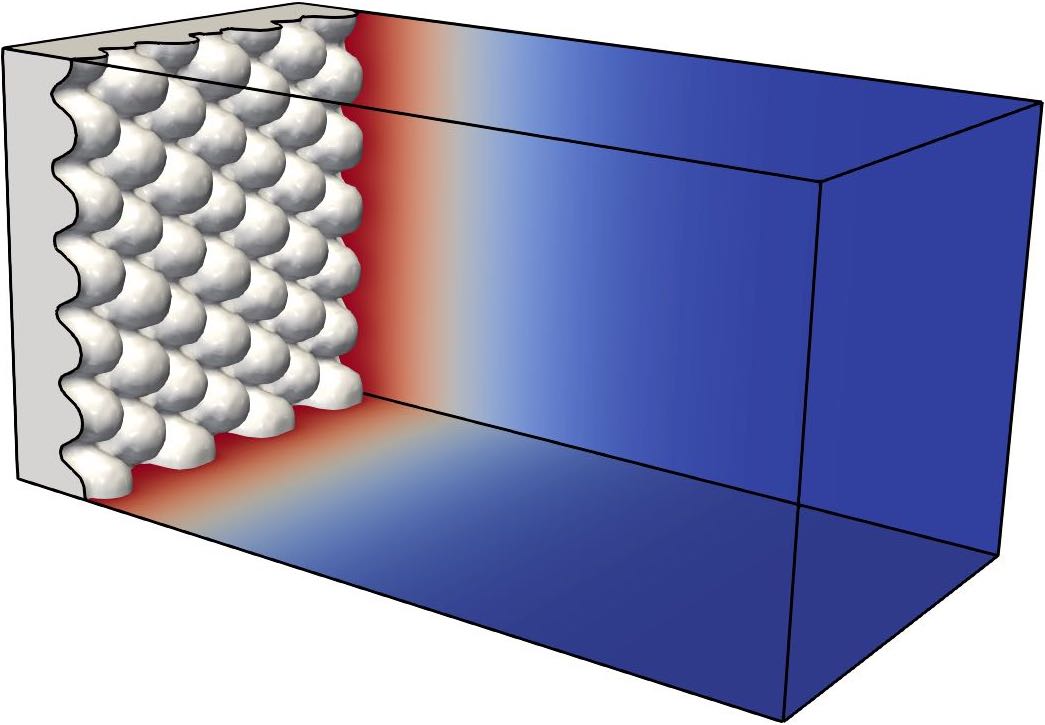}\vspace{2pt}
}%
\end{minipage} & {\small{}}%
\begin{minipage}[c]{2.7cm}%
{\small{}\vspace{2pt}
\includegraphics[width=0.8\textwidth]{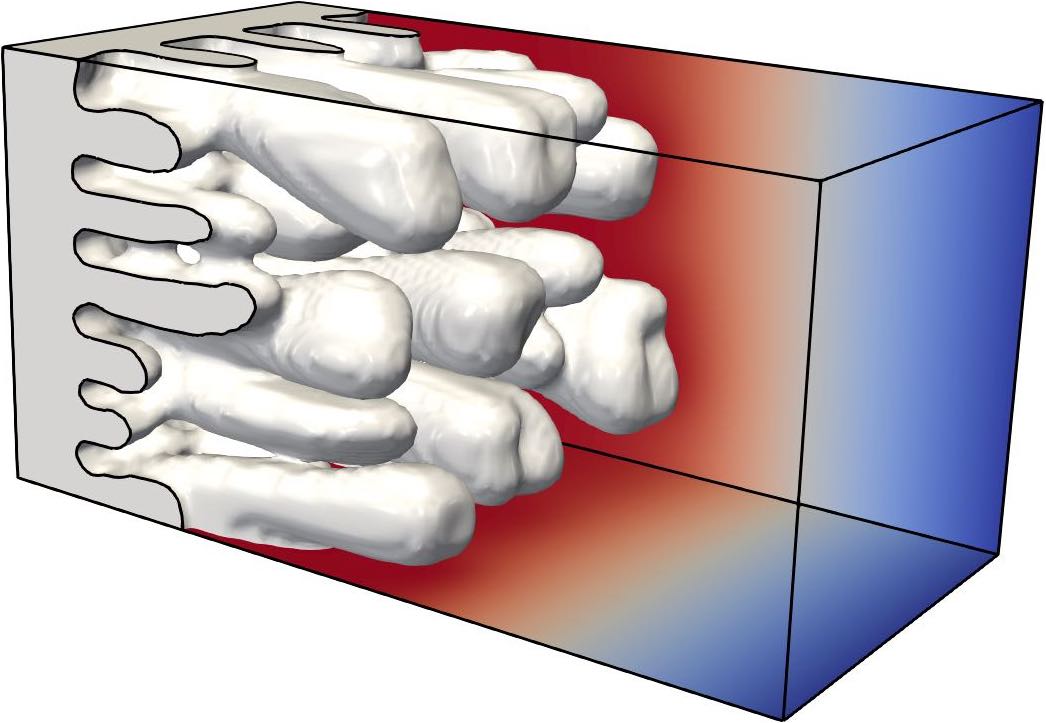}\vspace{2pt}
}%
\end{minipage} & {\small{}}%
\begin{minipage}[c]{2.7cm}%
{\small{}\vspace{2pt}
\includegraphics[width=0.8\textwidth]{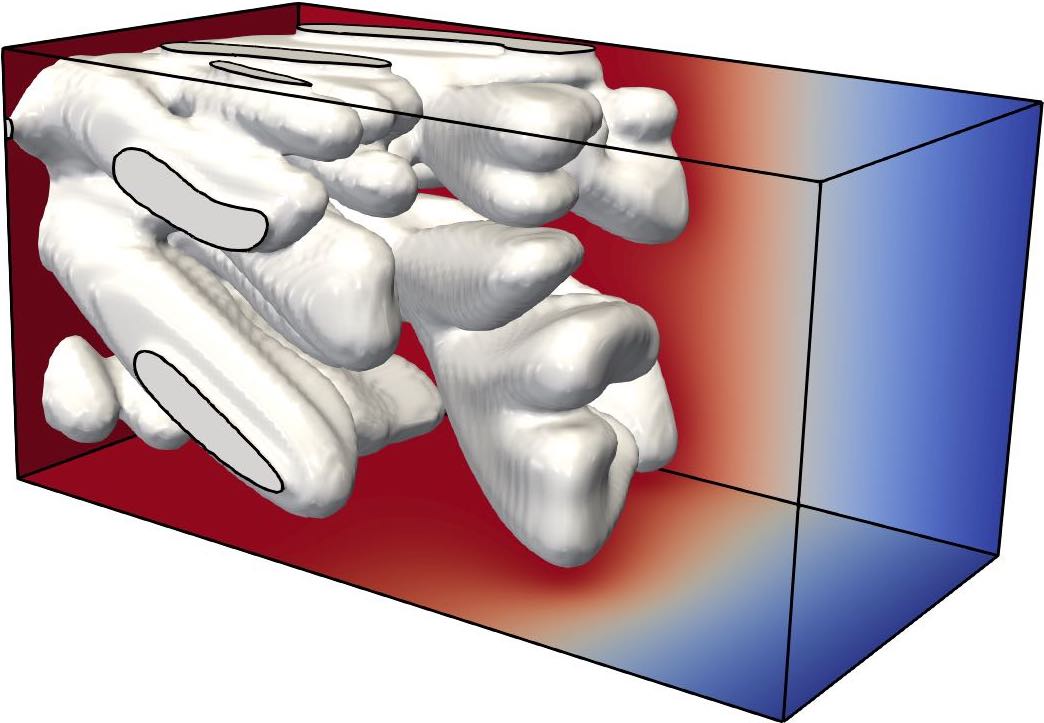}\vspace{2pt}
}%
\end{minipage}\tabularnewline
\cline{2-5} \cline{3-5} \cline{4-5} \cline{5-5} 
 & {\small{}TI$_{\left\langle 100\right\rangle }\!+\!\interp_{3}$}\\
{\small{}$\tilde{\width}=0.5$} & {\small{}}%
\begin{minipage}[c]{2.7cm}%
{\small{}\vspace{2pt}
\includegraphics[width=0.8\textwidth]{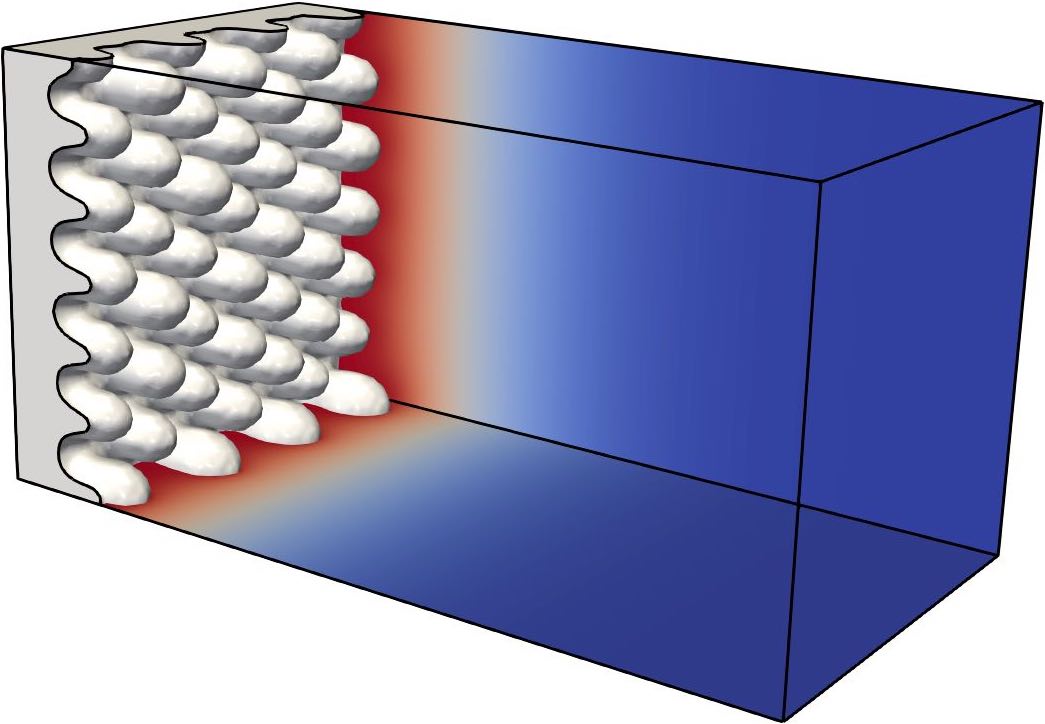}\vspace{2pt}
}%
\end{minipage} & {\small{}}%
\begin{minipage}[c]{2.7cm}%
{\small{}\vspace{2pt}
\includegraphics[width=0.8\textwidth]{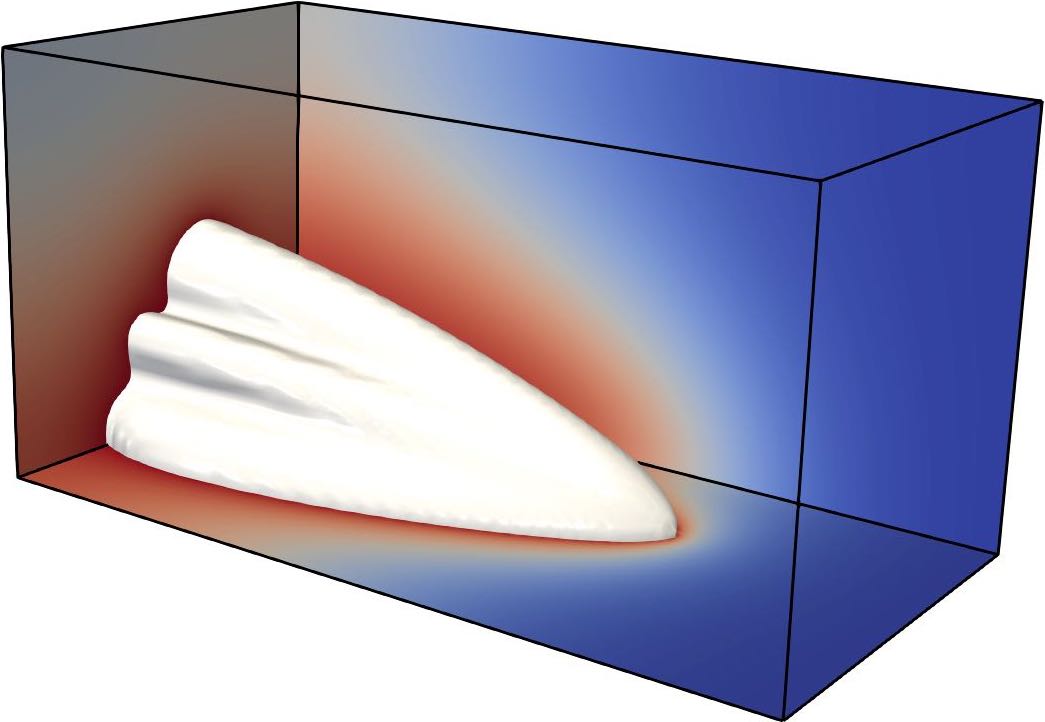}\vspace{2pt}
}%
\end{minipage} & {\small{}}%
\begin{minipage}[c]{2.7cm}%
{\small{}\vspace{2pt}
\includegraphics[width=0.8\textwidth]{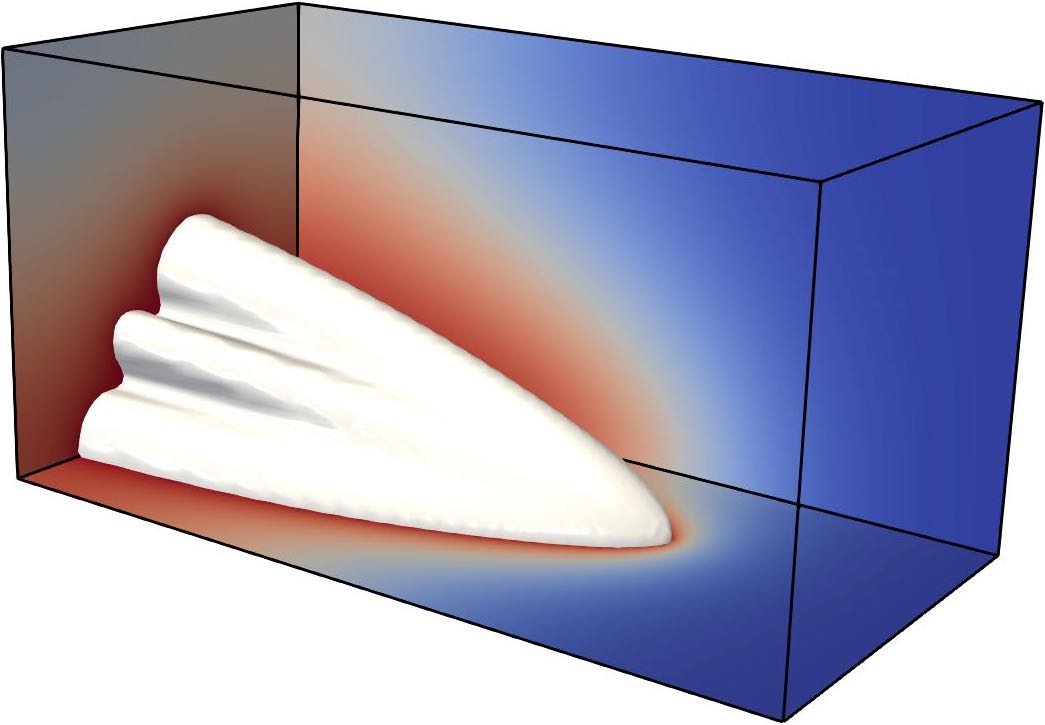}\vspace{2pt}
}%
\end{minipage}\tabularnewline
\cline{2-5} \cline{3-5} \cline{4-5} \cline{5-5} 
 & {\small{}CF$\!+\!\interp_{5}$}\\
{\small{}$\tilde{\width}=2.0$} & {\small{}}%
\begin{minipage}[c]{2.7cm}%
{\small{}\vspace{2pt}
\includegraphics[width=0.8\textwidth]{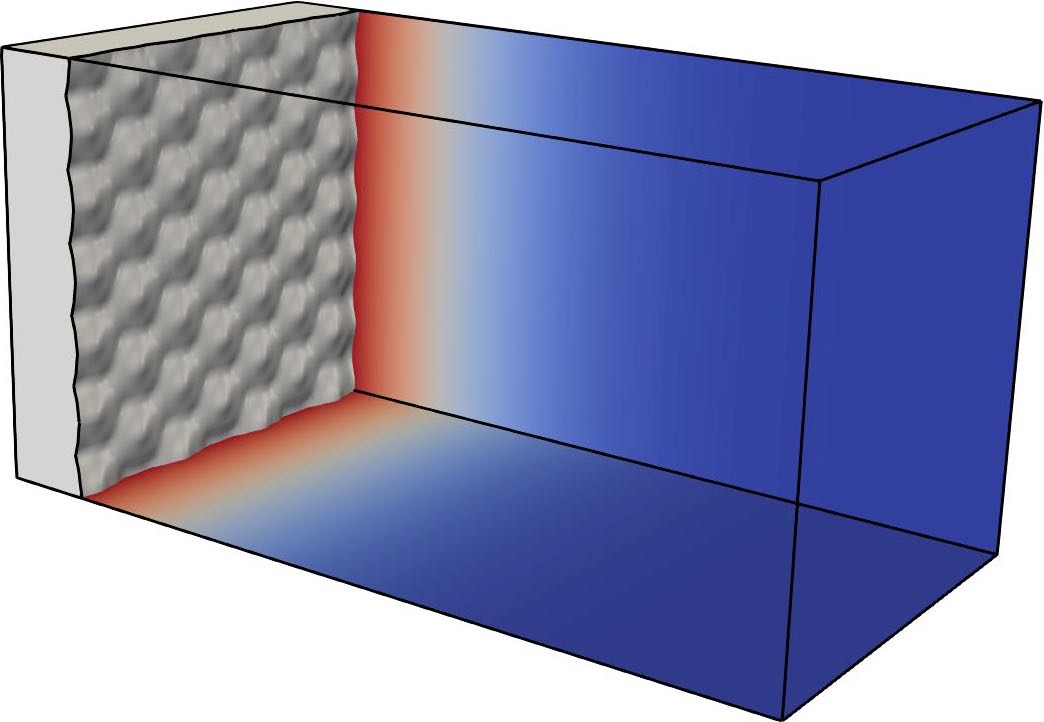}\vspace{2pt}
}%
\end{minipage} & {\small{}}%
\begin{minipage}[c]{2.7cm}%
{\small{}\vspace{2pt}
\includegraphics[width=0.8\textwidth]{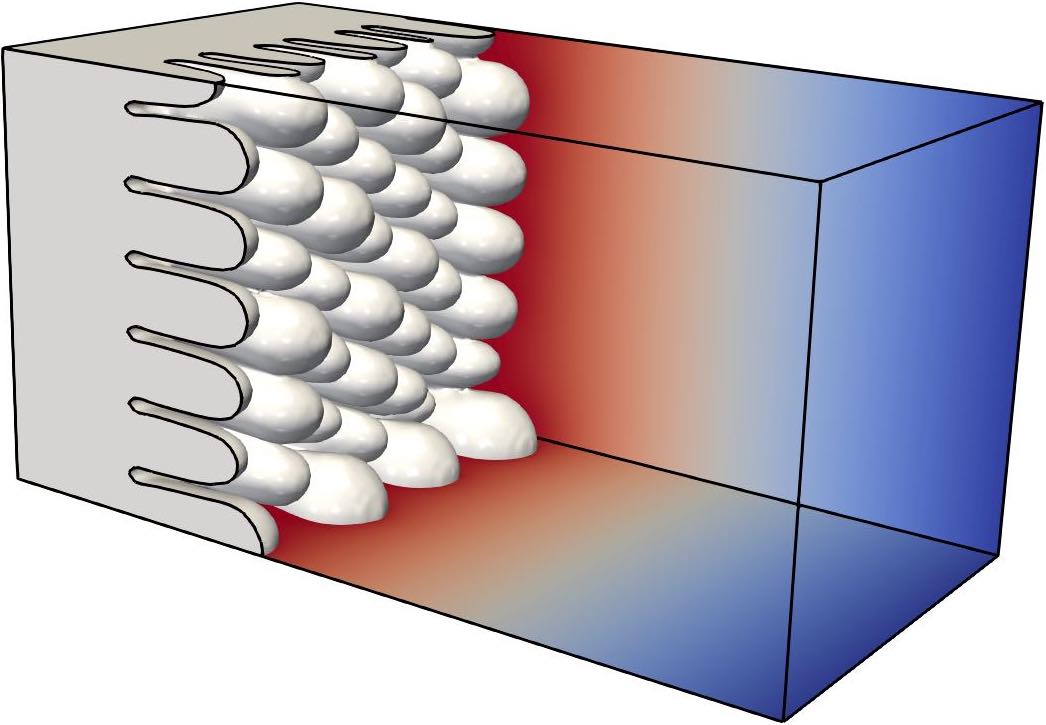}\vspace{2pt}
}%
\end{minipage} & {\small{}}%
\begin{minipage}[c]{2.7cm}%
{\small{}\vspace{2pt}
\includegraphics[width=0.8\textwidth]{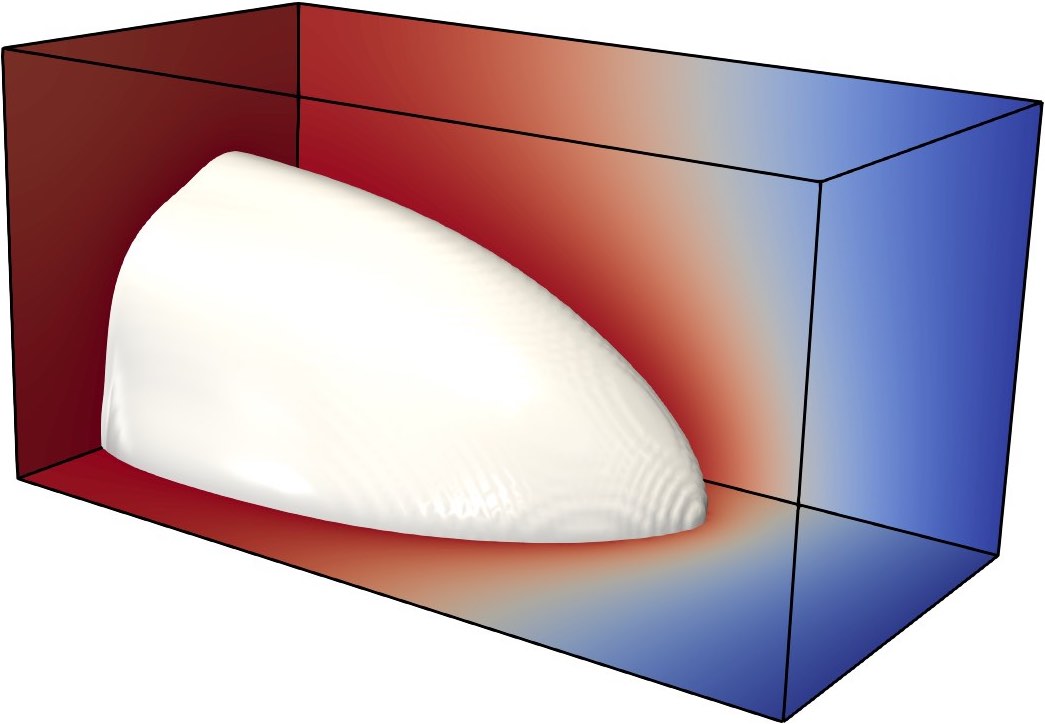}\vspace{2pt}
}%
\end{minipage}\tabularnewline
\hline 
\end{tabular}\caption{\label{fig:Demonstrator} Time series of phase-field simulations of
diffusion limited solidification using four different models: The
TI$_{\left\langle \mv{n}\right\rangle }\!+\!\interp_{3}$ model (i)
with and (ii) without regularization $\iFact$, (iii) the TI$_{\left\langle 100\right\rangle }\!+\!\interp_{3}$
model each with $\tilde{\width}\!=\!0.5$, and (iv) the CF$+\interp_{5}$-model
with $\tilde{\width}\!=\!2.0$. The temperature $\temperature$ is
visualized by the coloring and the phase-field is represented by the
$\phi\!=\!1/2-$contour. Further parameters: $\capLength/\dx\!=\!2\cdot10^{-3}$,
$D/\kin\!=\!5\!\cdot\!10^{-3}$, domain size $120\!\times\!60\!\times\!60$.}
\end{figure*}
For dimensionless undercooling temperatures $\temperature$ smaller
than unity, we obtain diffusion limited solidification. Four comparable
simulations are performed using four different phase-field models,
as shown in Fig.~\ref{fig:Demonstrator}. An animation showing the
full courses of all four simulations is provided in the supplementary
material. The simulations are started from the same initial state
at $\temperature=-0.3$. All boundaries are thermally insulating,
except for the boundary at the $\left[100\right]-$end of the simulation
domain on the right hand side, which is held at $\temperature_{\mathrm{max}}=-0.3$.
The initial quasi planar solid/liquid interface has small bumps at
regular intervals of $10\dx$. In the beginning, the interface develops
the Mullins-Sekerka instability \citep{MullinsSekerka1964}, since
the dimensionless capillary length is chosen to be sufficiently small
$\tilde{\capLength}\!=\!0.002$ ($\tilde{\mu}_{\mathrm{max}}\!=\!\temperature_{\mathrm{max}}/\tilde{\capLength}\!=\!150$).
As soon as the most advanced point of the solid/liquid interface exceeds
the fraction of $0.7$ of the simulation domain along the $\left[100\right]-$direction,
the whole system is shifted back by one grid point \citep{FleckBrenerSpatsch2010}.

In later stages, the disordered seaweed or dense-branching morphology
develops \citep{IhleMuel1994,BrenerHMKTemkinAbel1998,UtterBodenschatz2005},
if the residual grid anisotropy is sufficiently small. For super efficient
one-grid-point profile resolutions of $\tilde{\width}\!=\!0.5$, this
requires the local restoration of TI in the local interface normal
direction as well as the inclusion of the source term regularization
Eq.~(\ref{eq:SPFM-definition-iFact}), as shown in first row in Fig.~\ref{fig:Demonstrator}.
Without regularization the simulation shows a spurious dendritic selection
in the $\left\langle 110\right\rangle -$directions of the computational
grid, which originates from the inhomogeneous temperature release
via the source term in the diffusion equation. The simulations using
the TI$_{\left\langle 100\right\rangle }$- and CF- model show spurious
dendritic selection in the $\left\langle 100\right\rangle -$directions.
For the TI$_{\left\langle 100\right\rangle }$ model, the selection
originates from anisotropic interface kinetics \citep{Ihle2000,BragardKarmaLee2002},
which result from residual grid friction for interface orientations
that differ from the $\left\langle 100\right\rangle -$directions
\citep{FleckSchleifer2021}. In case of the CF-model, it results from
strong grid friction. 

\paragraph{Conclusion\label{subsec:Conclusion} }

A new sharp phase-field model is proposed: Instead of using global
grid dependent equilibrium potentials (\ref{eq:equilibrium-potential-TI}),
that restore the Translational Invariance (TI) for a finite amount
of fixed interface orientations, the newly proposed model restores
TI locally for the local interface normal direction $\mv{n}$. Furthermore,
we propose a source term regularization Eq.~(\ref{eq:SPFM-definition-iFact})
to effectively suppress spurious inhomogeneous temperature releasees
by diffuse interfaces as sharp as $\tilde{\width}\!=\!0.4$, see Fig.~\ref{fig:stationary-solidification-study}.
Compared to the conventional phase-field model with the resolution
limits $\tilde{\width}>2.0$ and $\tilde{\df}<1.0$, the sharp phase-field
model allows for super efficient quantitative simulations of stationary
solidification with phase-field profile resolutions of $\tilde{\width}\!=\!0.4$
and dimensionless driving forces up to $\tilde{\mu}\!=\!7200$! The
new sharp phase-field model with source term regularization (TI$_{\left\langle \mv{n}\right\rangle }\!+\!\interp_{3}\!+\!\iFact$)
provides extremely high degrees of isotropy. It provides the expected
isotropic seaweed or dense-branching morphology using extraordinary
efficient spatial resolutions: $\tilde{\width}\!=\!0.5$ and $\tilde{\mu}_{\mathrm{max}}\!=\!\temperature_{\mathrm{max}}/\tilde{\capLength}\!=\!150$!

\begin{acknowledgments}
We thank B.~Böttger and J.~Eiken from ACCESS, Aachen, Germany as
well as A.~Finel from ONERA, Ch\^atillon, France for fruit-full
discussions on this issue. The work is funded by the Deutsche Forschungsgemeinschaft
(DFG) -- 431968427.
\end{acknowledgments}

\noindent \noindent %
\begin{minipage}[t][1\totalheight][c]{0.2\columnwidth}%
\href{https://creativecommons.org/licenses/by/4.0/}{\includegraphics[width=1\textwidth]{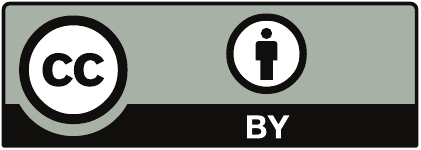}}%
\end{minipage} %
\begin{minipage}[t][1\totalheight][c]{0.78\columnwidth}%
\begin{spacing}{0.7}
{\footnotesize{}\copyright~2022. This version of the manuscript
is made available under the CC-BY 4.0 license \href{http://creativecommons.org/licenses/by/4.0/}{http://creativecommons.org/licenses/by/4.0/}}
\end{spacing}
\end{minipage}

\bibliographystyle{mynaturemag+title}
\bibliography{Literature-2Phase}

\noindent \newpage{}

\appendix

\appendixpagenumbering

\begin{widetext}
\begin{center}
\textbf{\large{}Supplementary material for: Sharp phase-field modeling
of isotropic solidification with a super efficient spatial resolution''}{\large\par}
\par\end{center}

In the supplementary material, we provide additional technical information
about the modeling and the simulations presented in the manuscript.

\end{widetext}

\subsection{Description of the supplementary animations}
\begin{enumerate}
\item \texttt{Supplementary\_\-material\_\-1\_\-stationary\_\-solidification\-.mpg:}
This movie is an animation showing the different simulations of stationary
solidification presented in Figure \ref{fig:stationary-solidification-study}.
It shows the time evolution of three different simulations of stationary
solidification, for a dimensionless undercooling temperature of $\temperature_{\mathrm{max.}}\!=\!-2$
and a dimensionless driving force of $\tilde{\df}_{\mathrm{min}}\!=\!100$.
\item \texttt{Supplementary\_\-material\_\-2\_\-diffusion\_\-limited\_\-solidification\-.mpg}:
This animation shows the four different simulations of diffusion limited
solidification presented in Figure \ref{fig:Demonstrator} using the
four different models: (i) TI$_{\left\langle \mv{n}\right\rangle }\!+\!\interp_{3}\!+\!\iFact$
with $\tilde{\width}\!=\!0.5$, (ii) TI$_{\left\langle \mv{n}\right\rangle }\!+\!\interp_{3}$
with $\tilde{\width}\!=\!0.5$, (iii) TI$_{\left\langle 100\right\rangle }\!+\!\interp_{3}$
with $\tilde{\width}\!=\!0.5$ and (iv) CF$+\!\interp_{5}$ with $\tilde{\width}\!=\!2.0$.
The temperature $\temperature$ is indicated by the coloring and the
phase-field is visualized via a $\phi\!=\!1/2-$contour plot. The
capillary length is $\capLength/\dx\!=\!0.002$, undercooling: $\temperature(\mv{x},0)\!=\!-0.3$,
kinetics: $D/\kin\!=\!0.01$. 
\end{enumerate}

\subsection{Stationary interface motion\label{subsec:Grid-friction}}

\begin{figure*}
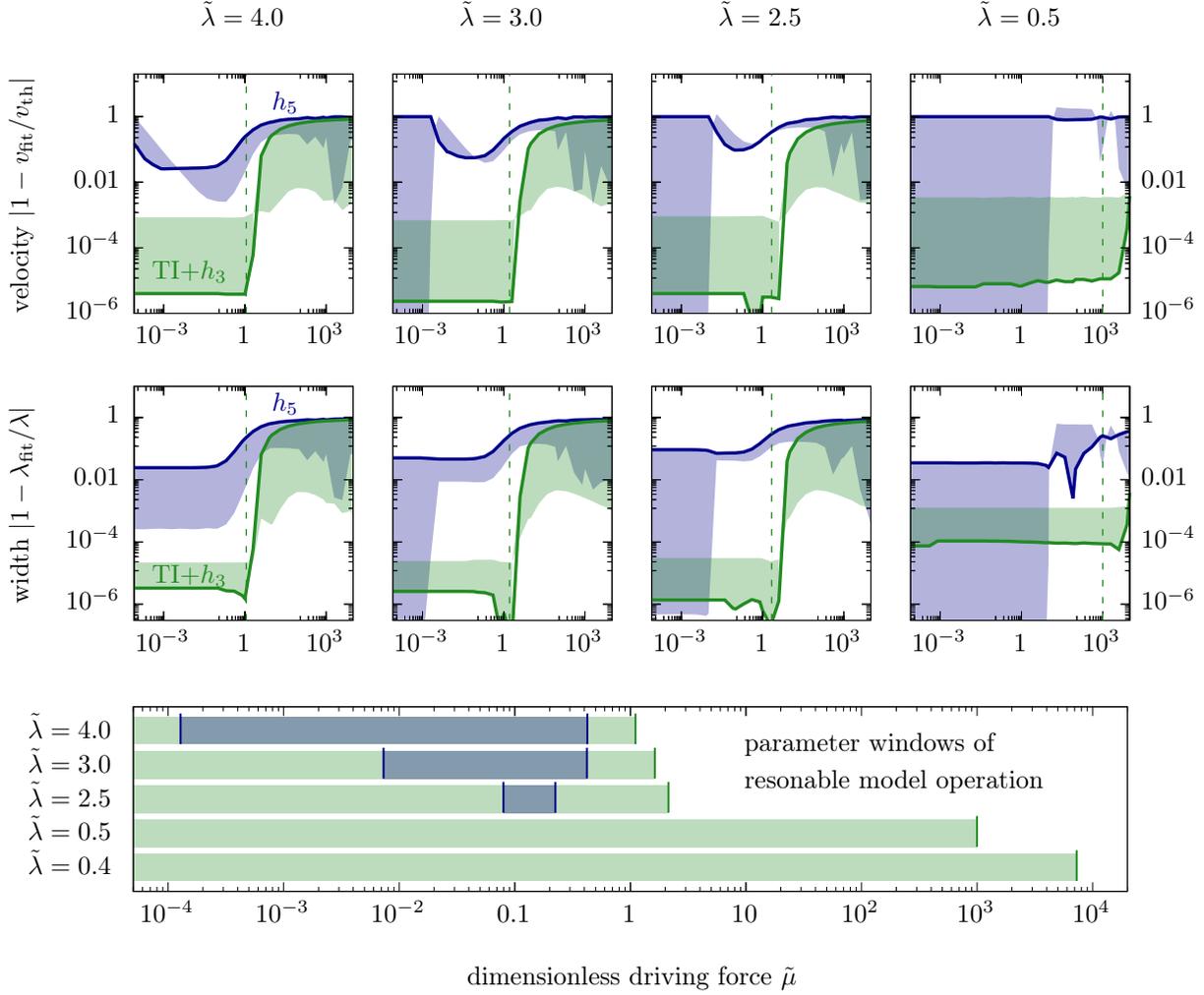

\begin{centering}
{\scriptsize{}}\begin{center}
\input{./VelocityVsdf-width-1.dtex}\hspace*{-1pt}\input{./VelocityVsdf-width4.dtex}\hspace*{0.5cm}\input{./VelocityVsdf-width3.dtex}\hspace*{0.5cm}\input{./VelocityVsdf-width2.dtex}\hspace*{0.5cm}\input{./VelocityVsdf-width0.dtex}\hspace*{-1pt}\input{./VelocityVsdf-width-2.dtex}
\par\end{center}

\begin{center}
\vspace{0.1cm}
\input{./WidthVsdf-width-1.dtex}\hspace*{-1pt}\input{./WidthVsdf-width4.dtex}\hspace*{0.5cm}\input{./WidthVsdf-width3.dtex}\hspace*{0.5cm}\input{./WidthVsdf-width2.dtex}\hspace*{0.5cm}\input{./WidthVsdf-width0.dtex}\hspace*{-1pt}\input{./WidthVsdf-width-2.dtex}
\par\end{center}

\noindent \begin{flushleft}
\hspace*{0.65cm}\input{./Evaluation_Parameter_Window_stationary-pf.dtex}\\
\par\end{flushleft}

\begin{center}
dimensionless driving force $\tilde{\df}$
\par\end{center}

\begin{center}
{\scriptsize{}\vspace{-35pt}
}{\scriptsize\par}
\par\end{center}{\scriptsize\par}
\par\end{centering}
\caption{\label{fig:Velocity-vs-driving-force}Errors plots of the stationary
interface velocity (top row) and the fitted interface width (middle
row) as a function of the dimensionless driving force $\tilde{\df}=\df\dx/\intEnergy$,
for different phase-field widths: $\tilde{\width}=\width/\dx=4.0,\;3.0,\;2.5,\;0.5$.
Two models are compared: (i) Continuum Field (CF) model with $\interp_{5}$
(blue) and (ii) the sharp phase-field model with Translational Invariance
(TI$+\interp_{3}$) (green). Solid lines denote the mean relative
errors and the oscillations are indicated as transparently colored
areas. The time resolution is $\kin\df\dt/(\intEnergy\dx)=1.6\cdot10^{-7}$.}
\end{figure*}
We consider the constantly driven stationary motion of a planar interface
in one dimension. In Fig.~\ref{fig:Velocity-vs-driving-force}, we
compare mean errors in the interface velocities and widths (solid
lines) as well as their relative oscillation amplitudes (colored areas)
for different models. As illustrated in the lower panel of Fig.~\ref{fig:stationary-solidification-study}a),
the colored areas start from the oscillation amplitude value and end
at the mean value. When the colored area is found above the mean value,
we have the ``healthy'' situation that the measured value oscillates
around the theoretic expectation. In contrast, colored areas below
the mean value denote the ``unhealthy'' case, when the theoretic
expectation is located outside the oscillation interval. While the
conventional Continuum Field (CF) model is subjected to pinning, the
sharp phase-field model allows for arbitrarily small driving forces.

The condition of phase stability demands the driving force to be small
enough to guarantee meta-stability of the high energy phase: The two
local minima of the potential energy density at $\phi=0,1$ have to
be separated by a maximum. The TI$+\interp_{3}$-model provides a
phase-field width dependent stability limit, which can be surprisingly
high. For instance, imposing the phase-field width $\tilde{\width}=0.4$,
then the limiting driving force is $\tilde{\left|\df\right|}_{\tilde{\width}=0.4}\lesssim7200$!
The theoretic stability limits for the different profile resolutions
$\tilde{\width}=\width/\dx=4.0,\;3.0,\;2.5,\;0.5$ have been indicated
by the vertical dashed green lines in Fig.\ref{fig:Velocity-vs-driving-force}.
These theoretical limits nicely reflect the behavior of the sharp
phase-field model. 

Switching the interpolation function changes the phase stability limits.
The most common choice for the interpolation function is $\interp_{\mathrm{5}}=\phi^{3}(10-15\phi+6\phi^{2})$.
The CF$+\interp_{5}$-model provides phase stability for infinitely
large driving forces! However, using interpolation functions other
than the natural one leads to altered nonequilibrium phase-field profiles.
The resulting deviation of fitted phase-field width $\width_{\mathrm{fit}}$
from the theoretic expectation $\width$ is plotted in the middle
row of Fig.~\ref{fig:Velocity-vs-driving-force}. The profile alternation
increases with increasing driving force. Increasingly stronger alternations
lead to increasingly stronger grid friction effects. Consider the
phase-field width $\tilde{\width}=3.0$ and $\tilde{\mu}=100$, then
the diffuse interface is compressed down to $22\%$ of its original
width. Grid friction drops the interface velocity down to about $5\%$
of the theoretic expectation. Thus, for large dimensionless driving
forces the CF-model $\interp_{5}$ is effectively limited by spurious
grid friction. In the lower part of Fig.~\ref{fig:Velocity-vs-driving-force},
we plot the parameter window of reasonable model operation. We define
the range of reasonable operation to end when the relative velocity
error exceeds $0.1$. 

\subsection{Translational Invariance of the ideal profile}

\label{subsec:Translational-invariance-test} Testing the Translational
Invariance (TI), we calculate the system integral over the equilibrium
condition (\ref{eq:discrete-equilibirum-formulation}). We consider
a discrete 3D system with a phase-field as represented by an array
of 64bit floating point numbers, each associated with a grid point
within the simple cubic numerical lattice of size $300\times1\times1$
(excluding the one stencil boundary halo). The phase-field values
are initialized according to the ideal profile function (\ref{eq:phase-field-tanh-profile-solution}),
such that the interface is sitting in the middle of the system. Then
the total grid friction forces are defined as the system integral
over (\ref{eq:discrete-equilibirum-formulation}). This integral value
may oscillate, when the ideal profile is moved in such a way that
the interface center $c_{n}$ passes several grid points. In Fig.~\ref{fig:Orientation-model},
we plot the oscillation amplitude $A$ of these forces for different
interface orientations. Large oscillation amplitudes indicate broken
Translational Invariance (TI). The conventional model without restoration
of TI is shown by the black curve. Restoring TI using constant grid
coupling parameters $\gridCoup_{\dir}\!\left(\mv{u}\right)$ based
on a globally constant unit vector $\mv{u}$, as proposed by Finel
et al., provides vanishing force oscillations for those interface
orientations, that match with one of the equivalent numerical lattice
directions $\left\langle \mv{u}\right\rangle $: When the grid coupling
parameters are, for instance, chosen based on the unit lattice vector
$\mv{u}$ parallel to the $\left[110\right]-$direction (TI$_{\left\langle 110\right\rangle }$,
see dark blue curve in Fig.~\ref{fig:Orientation-model}), then vanishing
force amplitudes are found for interface orientations with normal
vectors pointing in all the $\left\langle 110\right\rangle -$directions.
The vanishing force amplitudes are restricted to very sharp interface
orientation windows, as visible in Fig.~\ref{fig:Orientation-model}.
The new TI$_{\left\langle \mv{n}\right\rangle }$-model (green curve)
uses grid coupling parameters, that are determined by means of the
local interface normal direction. This leads to very small oscillation
amplitudes, regardless of the interface orientation. 

Translational Invariance (TI) of the ideal planar front solution can
also be tested with regard to oscillations in the total interface
energy. Then the system integral over the interface energy density
(\ref{eq:SPFM-Interface-energy-density-3D}) has to be evaluated instead.
However, in contrast to the forces, theoretically, only the total
interface energy, i.e.~the density integral over the full, infinite
profile, provides a TI value. The interface energy density alone does
not need to show this property. For a single direction $k$, the new
sharp phase-field formulation provides TI total interface energies
for arbitrarily oriented ideal phase-field profiles, as long as the
full profile function is evaluated in that direction.

\subsection{Construction of the models}

\label{subsec:Construction-of-models}Here, we explain how the different
models are constructed from the given finite difference equations.
A overview over all the different models is given in Tab.~\ref{tab:Overview-over-the-models}.
The models differ by different choices for the equilibrium potentials
$g_{\dir}(\phi)$ and for the interpolation function $\interp(\phi)$.
Further, the source term regularization factor $\iFact(\phi)$ can
be either imposed or otherwise set to unity. All models are separately
calibrated. Thus, the imposed calibration parameters, $C_{\intEnergy}$,
$\ponderation_{\nbs}$, can be different for the different models.
The Continuum Field (CF) model is obtained in the limit $\lim_{\left|\mv{u}_{\dir}\right|\rightarrow0}$.
In this limit the equilibrium potentials (\ref{eq:equilibrium-potential-TI})
converge to the classical quartic double-well potential. For the CF-model,
we impose the equilibrium potentials $g_{\dir}^{\infty}\!=\!\bar{\nu}8\phi^{2}(1\!-\!\phi)^{2}$,
where the multiplicity correction $\bar{\nu}\!=\!1/3$ equilibrates
for the overweighting by the sum in the equilibrium potentials within
each neighboring shell $\nbs$.

\begin{widetext}
\begin{table}[H]
\centering{}\caption{\label{tab:Overview-over-the-models}The construction of the different
models.}
\begin{tabular*}{1\textwidth}{@{\extracolsep{\fill}}lclclcccl}
\hline 
\noalign{\vskip2pt}
model & ~ & equilibrium potential & ~ & interpolation function & ~ & regularization & ~ & \multicolumn{1}{l}{calibration}\tabularnewline[2pt]
\hline 
\noalign{\vskip2pt}
CF$+\!\interp_{5}$ &  & \multirow{1}{*}{$g_{\dir}^{\infty}\!=\!\bar{\nu}8\phi^{2}(1\!-\!\phi)^{2}$} &  & $\interp_{\mathrm{5}}\!=\!\phi^{3}(10\!-\!15\phi\!+\!6\phi^{2})$ &  & -- &  & \multirow{1}{*}{$C_{\intEnergy}^{\mathrm{CF}}$, $\ponderation_{\nbs}^{\mathrm{CF}}$}\tabularnewline
TI$_{\left\langle 100\right\rangle }\!+\!\interp_{3}$ &  & $g_{\dir}$: Eq.~(\ref{eq:equilibrium-potential-TI}), $\gridCoup_{\dir}(\mv{u}_{\left\langle 100\right\rangle })$ &  & $\interp_{3}\!=\!\phi^{2}(3\!-\!2\phi)$ &  & -- &  & $C_{\intEnergy}$, $\ponderation_{\nbs}^{\mathrm{TI}_{\left\langle 100\right\rangle }}$\tabularnewline
TI$_{\left\langle \mv{n}\right\rangle }\!+\!\interp_{3}$ &  & $g_{\dir}$: Eq.~(\ref{eq:equilibrium-potential-TI}), $\gridCoup_{\dir}\left(\mv{n}\right)$ &  & $\interp_{3}\!=\!\phi^{2}(3\!-\!2\phi)$ &  & -- &  & $C_{\intEnergy}$, $\ponderation_{\nbs}^{\mathrm{TI}_{\left\langle \mv{n}\right\rangle }}$\tabularnewline
TI$_{\left\langle \mv{n}\right\rangle }\!+\!\interp_{3}\!+\!\iFact$ &  & $g_{\dir}$: Eq.~(\ref{eq:equilibrium-potential-TI}), $\gridCoup_{\dir}\left(\mv{n}\right)$ &  & $\interp_{3}\!=\!\phi^{2}(3\!-\!2\phi)$ &  & $\iFact:$ Eq.~(\ref{eq:SPFM-definition-iFact}) &  & $C_{\intEnergy}$, $\ponderation_{\nbs}^{\mathrm{TI}_{\left\langle \mv{n}\right\rangle }}$,
$C_{\iFact}$\tabularnewline[2pt]
\hline 
\end{tabular*}
\end{table}

\end{widetext}

Translational Invariance (TI) is obtained when the new equilibrium
potentials Eqs.~(\ref{eq:equilibrium-potential-TI}) are imposed
in conjunction with the natural interpolation function $\interp_{3}$.
When all the grid coupling parameters $\gridCoup_{k}$ in the equilibrium
potentials are set as fixed, based on the globally fixed lattice vector
$\mv{u}\!=\!\left[100\right]$, then TI is restored for all equivalent
$\left\langle 100\right\rangle -$directions of the computational
grid. This model is denoted as TI$_{\left\langle 100\right\rangle }+\interp_{3}$.
A combination of the new equilibrium potentials with the other interpolation
function is not useful, because the nonequilibrium phase-field profile
alternation destroys the carefully restored TI again. In case of the
TI$_{\left\langle \mv{n}\right\rangle }$-models, the locally calculated
and length corrected grid coupling parameters $\gridCoup_{k}\left(\mv{n}\right)$
(see Eq.~(\ref{eq:grid-coupling-calculation-scheme}) ff.) are used
in the equilibrium potentials $g_{\dir}(\phi)$ Eq.~(\ref{eq:equilibrium-potential-TI}).

\subsection{Additional information on the modeling}

The phase-field equation of motion is given as 
\begin{align}
\partial_{t}\phi_{\grid}= & \frac{2\kin}{3C_{\intEnergy}}\sum_{\nbs,\dir}\ponderation_{\nbs}\nu_{\nbs}\left(\partial_{\dir}^{-}\left(\partial_{\dir}^{+}\phi_{\grid}\right)-\frac{1}{\width^{2}}\partial_{\phi}g_{\dir}(\phi_{\grid})\right)\nonumber \\
 & -\frac{2\kin}{3\width\intEnergy}\df\;\partial_{\phi}\interp(\phi_{\grid}),\label{eq:Allen-Cahn-Phase-field-equation}
\end{align}
where $\kin$ is a kinetic coefficient comparable to a diffusion coefficient
with dimension $\left[\kin\right]=\mathrm{m}^{2}\mathrm{s}^{-1}$.
 We know that the phase-field equation promotes solution of the form
of Eq.~(\ref{eq:phase-field-tanh-profile-solution}).  The hyperbolic
tangent function provides the following addition property, 
\begin{align}
\tanh\left(p\pm q\right) & =\frac{\tanh\left(p\right)\pm\tanh\left(q\right)}{1\pm\tanh\left(p\right)\tanh\left(q\right)}.
\end{align}
This property can be reformulated in terms of the phase-field profile
function, and we obtain the relation 
\begin{align}
\phi_{\grid\pm\gvec} & =\frac{\left(1\pm\gridCoup_{\dir}\right)\phi_{\grid}}{1\pm\left(2\phi_{\grid}-1\right)\gridCoup_{\dir}},\label{eq:profile_property}
\end{align}
 where the grid coupling parameter $\gridCoup_{\dir}$ has been introduced
as $\gridCoup_{\dir}=\tanh\left(2\gvec\cdot\mv{n}/\width\right).$The
equilibrium condition (\ref{eq:discrete-equilibirum-formulation})
holds, if all 1D $\dir-$components are simultaneously satisfied.
The individual $\dir-$component can be satisfied at any real time
during the propagation of the interface by using the addition property
of the hyperbolic tangent profile. Inserting (\ref{eq:profile_property})
into the $\dir-$th component of the equilibrium condition (\ref{eq:discrete-equilibirum-formulation})
yields  
\begin{align}
\partial_{\phi}g_{\dir}(\phi) & =\gridCoup_{\dir}^{2}\frac{\width^{2}}{\gvec^{2}}\frac{4\phi\left(1-\phi\right)\left(1-2\phi\right)}{1-\gridCoup_{\dir}^{2}\left(1-2\phi\right)^{2}}.\label{eq:derive-equilibirum-protential-TI}
\end{align}

\end{document}